\documentclass[10pt,twocolumn,twoside]{IEEEtran} 
   \usepackage[pdftex]{graphicx}
  \DeclareGraphicsExtensions{.pdf,.jpeg,.png}
\usepackage[]{units}
\usepackage[T1]{fontenc} 
\usepackage{amsmath}
\usepackage{amssymb}
\usepackage{bm} 
\usepackage{eurosym}
\usepackage[super]{nth} 
\usepackage{multirow}
\usepackage{color}
\usepackage[caption=false,font=footnotesize]{subfig}
\usepackage{enumitem}
\usepackage{cite}

\usepackage{longtable} 
\usepackage{eurosym}
\usepackage{comment}
\usepackage{float}

\usepackage[export]{adjustbox}
\usepackage[subpreambles=true]{standalone}
\usepackage{algorithm,caption}
\usepackage{algpseudocode} 
\usepackage{multicol}
\usepackage{tabularx}
\usepackage[dvipsnames]{xcolor}
\begin{document}
\raggedbottom
%
\title{A Novel Decentralized Inverter Control Algorithm for Loss Minimization and LVRT Improvement}
%
%
%

\author{Ilgiz~Murzakhanov,~\IEEEmembership{Graduate~Student~Member,~IEEE,}
        Gururaj~Mirle~Vishwanath,~\IEEEmembership{Member,~IEEE,}
        Kasi~Vemalaiah,~\IEEEmembership{Graduate~Student~Member,~IEEE,}
        Garima~Prashal,~\IEEEmembership{Graduate~Student~Member,~IEEE,}
        Spyros~Chatzivasileiadis,~\IEEEmembership{Senior~Member,~IEEE,}
        Narayana~Prasad~Padhy,~\IEEEmembership{Senior~Member,~IEEE}
\thanks{This work is supported by the project entitled Indo-Danish collaboration for data-driven control and optimization for a highly Efficient Distribution Grid (ID-EDGe) funded by Innovation Fund Denmark, Grant Agreement No. 8127-00017B and Department of Science and Technology (DST), India, Grant No: DST-1390-EED. }
\thanks{I. Murzakhanov and S. Chatzivasileiadis are with the Department of Wind and Energy Systems, Technical University of Denmark (DTU), Kgs. Lyngby, Denmark. E-mail: \{ilgmu, spchatz\}@dtu.dk.}
\thanks{Gururaj M.V. is with the Department of Electrical Engineering, Indian Institute of Technology Kanpur, India. Email: gururajmv@iitk.ac.in.}
\thanks{K. Vemalaiah, G. Prashal, and N.P. Padhy are with the Department of Electrical Engineering, Indian Institute of Technology Roorkee, India. E-mail: \{kasi\textunderscore v, garima\textunderscore p, nppadhy\}@ee.iitr.ac.in.}}
\maketitle


\begin{abstract}
Algorithms that adjust the reactive power injection of converter-connected RES to minimize losses may compromise the converters’ fault-ride-through capability. This can become crucial for the reliable operation of the distribution grids, as they could lose valuable resources to support grid voltage at the time they need them the most. This paper explores how two novel loss-minimizing algorithms can both achieve high reduction of the system losses during normal operation and remain connected to support the voltage during faults. The algorithms we propose are decentralized and model-free: they require no communication and no knowledge of the grid topology or the grid location of the converters. Using local information, they control the reactive power injection to minimize the system losses. In this paper, we extend these algorithms to ensure the low voltage ride through (LVRT) capability of the converters, and we integrate them with state-of-the-art Wavelet-CNN-LSTM RES forecasting methods that enhance their performance. We perform extensive simulations on the real-time digital simulation (RTDS) platform, where we demonstrate that the algorithms we propose can achieve a substantial decrease in power losses while remaining compliant with the grid codes for LVRT makes them suitable for the implementation across the distribution system. 
\end{abstract}


%
\IEEEpeerreviewmaketitle

\section{Introduction}
%
%
%
%
\IEEEPARstart{E}{nhanced} environmental awareness and the concerns for the security of supply of energy sources lead to plans for drastically increasing penetration of renewable energy sources (RESs) to the existing power system \cite{gur1}. Amongst the existing RESs, photovoltaic (PV) and wind systems are the most popular when it comes to investments in new electric power generation. Being to some extent complementary in nature, the combination of solar and wind systems can increase their overall reliability \cite{gur2}. Among the available wind generator types, the majority of the existing wind turbine installations are of the semi-variable type of doubly fed induction generators (DFIG). DFIG used to be the most popular due to its cost benefits obtained with the utilization of partial rated converters and also its capability of extracting maximum power from the wind turbine (WT) over a given range of wind speeds \cite{gur5}.  Newer wind farms mostly install type-IV wind turbines, where the wind electromechanical system is connected through a full-rated AC/DC/AC converter to the rest of the system. The proposed algorithms can also provide loss minimization service in wind turbines of type-IV.

\subsection{Loss Minimization Techniques and Their Demerits}
Control of PV inverters may have several goals, including minimization of active power losses and improvement of voltage profile. All possible approaches explored during this literature review can be classified by the presence or absence of a central coordinator and by the need or not for communication between agents. Control schemes requiring a central coordinator can potentially achieve the best possible performance, but they often require frequent exchanges of large amounts of data, reliable communication infrastructure and some sort of reliable timestamping to achieve synchronized measurements and ensure an appropriate coordinated control response. Coordinated control schemes can become challenging to implement at scale, considering the control of millions of converter based resources. On the other hand, distributed or local methods do not require central coordination and often have much lower communication needs. Although they do not always achieve the performance centralized methods do, they perform better at scale. In this work, we propose algorithms that do not need a central coordinator and require no communication. Considering the abundance of possible approaches, we limit the scope of this literature review to decentralized methods existing in the literature. Ref. \cite{Ahn2013} proposes a model-free decentralized algorithm for minimizing power losses. The performance of the introduced two-level algorithm appears to be highly dependent on the communication network: the version without communication does not reach the minimum loss condition, has slower convergence and fluctuating performance. In \cite{Mousa2019}, the authors have presented an affinely adjustable robust counterpart (AARC) approach for improving the voltage profile. The approach requires though information about the line parameters in the system. A decentralized impedance-based adaptive droop method for power loss reduction has been presented in \cite{Oureilidis2016a}. As the droop coefficients depend on the microgrid impedance, information on the electrical parameter of the connection lines is needed in this approach as well. Ref. \cite{Ghosh2014} develops a droop algorithm for voltage control by reactive power injections from PV inverters, but the proposed droop control is based on heuristic rules. As a result, there is no guarantee of proper execution of the algorithm for the power system with an unknown topology. In \cite{Kundu2013}, the local strategy algorithm requires a parameter, that is computed as the reactance to the resistance ratio of distribution lines, for minimizing voltage deviations and line losses. Similarly, the control of distributed PV generators in \cite{Jabr2018} exploits information of the network nodal admittance matrix. In \cite{Weckx2016}, the authors propose designing an optimal $Q(P)$ curve that keeps the voltage within the limits. The drawback of the proposed approach is the requirement for extensive data of voltage and generation of RESs. Additionally, the method can arrive at the state with higher power losses compared to the scenario without reactive power control. In contrast to all considered methods, our solution has a proven mathematical guarantee for the minimization of power losses for any distribution grid without requiring the communication of any non-local information.

\subsection{LVRT and Grid Codes}
As discussed above, many works propose loss minimization techniques; these usually require the adjustment of the reactive power setpoints of the converter-based resources. Reactive power reserves, however, are necessary to also ensure the fault-ride-through (FRT) capability of these devices. Especially when it comes to RES, the availability of reactive power reserves for FRT can become crucial, as it ensures that RES remain interconnected during temporary fault and support the reactive power avoiding voltage collapse \cite{gur14}. Higher reactive power reserve in the system during normal conditions results in greater reactive power support during the fault which further leads to improvement in the voltage stability as it reduces the post fault voltage recovery time \cite{gurIm33}. On the contrary, the higher the reactive power reserve kept during the normal conditions (for example, $100\%$ of the available capacity), the lower is the capability of the converter-based resources to assist in loss minimization ($0\%$ available for loss minimization). It becomes clear that there is a trade-off between the LVRT capability of converter-based resources and their ability to adjust their reactive power setpoints for loss minimization. So far, existing approaches in the literature have not considered how loss minimizing algorithms use the reactive power reserves and if this affects the LVRT capabilities of the resources. The work in this paper considers this interaction and moves a step further: it develops a novel way of determining the optimal reactive power setting of the inverter during normal working conditions which varies along with the active power generation of the renewables. Combining this with RES forecasting techniques (see Section~\ref{sec:litreview_forecasting}), we are able to determine the most appropriate dynamic reactive power setting, assisting in loss minimization and ensuring LVRT despite the varying reactive power reserves.

\begin{figure}
    \centering
    \includegraphics[width=0.9\linewidth]{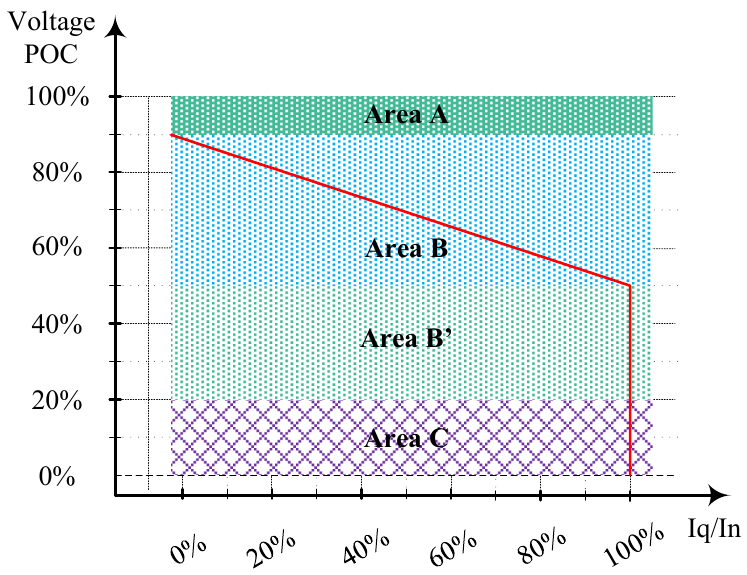}
    \caption{Requirements for reactive power support to the grid by wind turbines, according to the Danish grid code\cite{6672149}.}
    \label{fig:grid_code_lvrt}
\end{figure}

In this paper, we implement the Danish grid code requirements. Fig.~\ref{fig:grid_code_lvrt} presents the requirements for reactive power support to the grid by the Wind Power Plants, as stipulated by the Danish grid code. $I_q/I_n$ on the x-axis denotes the ratio of the reactive current delivered/absorbed by a WT ($I_q$) to the maximum continuous current that a WT is designed to deliver ($I_n$); the y-axis shows the normal operation voltage at the point of connection (POC). Area A corresponds to normal operation when a distributed generation (DG) unit must remain connected to the grid. On the contrary, area C permits disconnecting the DG. In areas B and B', the DG must stay connected to the grid and provide maximum voltage support by supplying an added amount of controlled reactive current. For voltage magnitudes within $[0.5; 0.9]$ p.u. the amount of reactive current is defined by the slope of the red line in area B of Fig.~\ref{fig:grid_code_lvrt}. Finally, if voltage drops in the range of $[0.2; 0.5]$ p.u., the DG must inject reactive current equal to $100\%$ of its capacity, according to area B' of Fig.~\ref{fig:grid_code_lvrt}. Voltage support in areas B and B' should continue until the voltage returns to normal operation, i.e. in area A. 

\subsection{Existing Forecasting Techniques}
\label{sec:litreview_forecasting}
The forecasting of solar and wind power time series is a complex regression task due to their inherent stochastic characteristics. Statistical forecasting, numerical weather predictions (NWP), and artificial intelligence are the main categories of the renewable forecasting techniques \cite{sayeed2021deep}. The mathematical relationship involved in the statistical method restrains its efficacy only for short-term forecasting applications. On the other hand, NWP involves forecasting weather parameters which is a computationally demanding task, thus it has limited utility for long and medium-term forecasting applications. Recent advancements in the field of neural networks have demonstrated their exceptional ability in prediction problems. Neural networks can be used for short, medium, and long-term forecasting because of their ability to determine the complex nonlinear relationship between renewable power and different weather parameters. From the plethora of forecasting methods based on neural networks, we choose Wavelet-CNN-LSTM for the following reasons. First, wavelet transform has been successfully used to decompose low- and high-frequency components of stochastic signals, such as solar irradiance and wind speed \cite{performance}. Second, CNN (convolution neural networks) can gradually transform local features into global characteristics and, as a result, effectively reduce training parameters and training time \cite{cnn-lstm}. Third, LSTM (long short-term memory) networks are well-known for accurately capturing long-term dependencies of time series data \cite{cnn-lstm}. In Section IV-A, we benchmark the chosen Wavelet-CNN-LSTM with LSTM, CNN-LSTM, and ARIMA (autoregressive moving average) \cite{arima}.

\subsection{Main Contributions}
This paper has the following contributions:
\begin{itemize}
    \item We extend the local load (LLMA) and local flow (LFMA) measuring algorithms for minimizing the active power losses, first presented in \cite{arxiv_external}, to examine if they comply with the low voltage ride through (LVRT) grid codes of Denmark.
    \item We validate the performance of the proposed techniques for various test scenarios on the IEEE 33-bus system using the real-time digital simulation (RTDS) platform. According to the numerical results, LLMA and LFMA maintain the FRT capabilities of the inverters, while they manage to significantly decrease power losses.
    \item To  further extend the loss minimization capabilities of LLMA and LFMA, we introduce a novel algorithm for dynamic adjustment of reactive power setpoint depending on the predicted active power output of RES. 
    \color{black}
    The novelty of the proposed local algorithms consists in open-loop setup since the algorithms do not require measuring voltage, but only reactive power. As a result, the local algorithms are guaranteed to converge. An additional novelty is introduced by dynamic adjustment of reactive power setpoint, when by utilizing the forecast data additional power savings are made possible.  
    \color{black}
    Namely, for forecasting active power generation of PV and WT, we implement the state-of-the-art Wavelet-CNN-LSTM. Notably, the inclusion of the adaptive control and Wavelet-CNN-LSTM enhances the ability of the converters to remain connected during faults. 
    \item Finally, through full-day simulations, we demonstrate the economic savings due to the proposed algorithms if the inverters are kept grid-connected at night hours. 
\end{itemize}

\subsection{Outline}
The rest of the paper is organized as follows. Section II describes implemented forecasting techniques. Section III introduces proposed loss minimization algorithms. In Section IV, we present the numerical results of the RTDS simulations. Section V concludes the work. 

\section{PV and Wind Forecasting}
In this section, we describe the used dataset, the pre-processing steps, and the application of the implemented forecasting techniques.
\subsection{Data Site Description and Preprocessing}
The real-time solar and wind dataset recorded at SYSLAB, Technical University of Denmark, is used to validate the forecasting model\cite{syslab}. The dataset model consists of 2 wind turbines with a capacity of 11 kW and 10 kW and 3 PV plants with 10 kW, 10 kW, and 7 kW. The data used in this work is recorded at the frequency of \textcolor{black}{5}-minute, from January 1, 2019 to December 31, 2019. The input dataset for PV forecasting is the temperature $(^\circ C )$, humidity $(\%)$, wind speed $(m/s)$, and solar irradiance $(W/m^2)$. All inputs are the same for wind forecasting, except instead of solar irradiance, wind direction (angle) is considered based on correlation analysis.
The considered one-year solar and wind datasets have $0.62\%$ and $0.67\%$ missing values, respectively. Missing values are common in real-world datasets and appear due to failure to record measurements at some time intervals. We handle the missing values using the linear interpolation technique. The input parameters are normalized between 0 and 1 using MinMaxScaler function in Python. The normalized one year datasets of PV and wind are used for training and testing in the ratio of 9:1, respectively. In this work, we focus on very short term forecasting i.e. one-minute ahead forecasting.

\subsection{Wavelet-CNN-LSTM Framework}

\subsubsection{Wavelet Transform}
Solar and wind data contain non-periodic oscillations and ramps due to their stochastic behavior. These oscillations contain various frequency components, which could be caused by abrupt changes in weather circumstances or sensor malfunction. The idea is to capture these spikes in their respective frequency domain and make the predictor learn each coefficient in its own frequency domain. The discrete wavelet transform (DWT) technique is used to capture these spikes in their respective frequency domains. DWT decomposes the solar and wind power signal at several stages by downsampling it using high and low pass filters. DWT downsamples the signal in numerous phases, utilizing high and low pass filters to deconstruct it. In this work, a three-stage decomposition is used by creating high-frequency components (D1, D2, and D3) as well as a low-frequency component (A3) \cite{performance}. The difference between D1, D2, and D3 is that D1 captures the biggest amount of the noise in input, and D2 captures less than D1 but more than D3. The number of high-frequency components is defined for each problem individually, and three high-frequency components work the best for our forecasting task. 

\subsubsection{CNN}

 1-D convolution neural network is used to capture the spatial relation present in the solar and wind datasets. The two layers of CNN are used with a filter size of 16 and  kernel size of 4 to extract the features from the input matrix. The sparse characteristic of CNN requires less memory, which lessens the computational burden and memory space utilization \cite{cnn1}. The element-wise convolution of the input matrix is calculated using the 1-D filter.
The ReLU activation function is used here as it avoids the vanishing gradient problem and provides faster computation than tanh and sigmoid activation functions. A max-pooling layer of  $2 \times 2$ size is used to downsample the features, which further improves the computational efficiency. 

\subsubsection{LSTM}
The vectorized output of CNN is given as an input to the LSTM layer to learn the temporal relation. The memory cell of LSTM consists of three gates: input, output, and forget gate which is responsible for maintaining the sequence of the network. The forget gate controls the information of past cells, and the input gate decides what information needs to be preserved in the internal state of a memory cell. After updating the internal state, the output gate provides the LSTM's output. More details can be found in \cite{cnndetail}. 

The architecture of LSTM incorporates three LSTM layers each with 150 neurons. The dropout layers are provided with a $0.5$ probability to handle the overfitting problem of CNN-LSTM which is caused due to increased parameters. The dense layer of the CNN-LSTM model with 50 neurons and softmax activation function provides the predicted value of renewable power.

\begin{figure}
    \centering
    \includegraphics[width=0.9\linewidth]{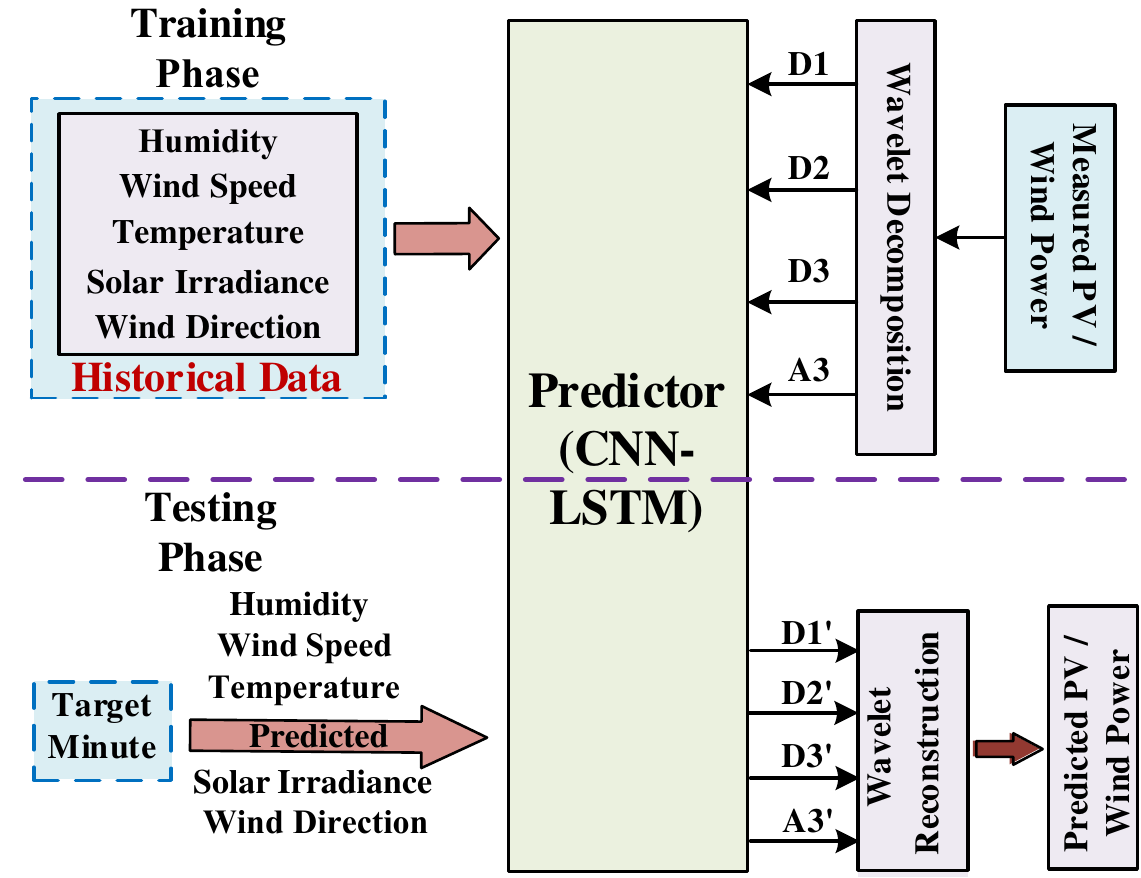}
    \caption{Renewable energy forecasting framework.}
    \label{fig:forecast}
\end{figure}

In this work, the same training process is followed for both PV and wind forecasting to make the model learn the relationship between weather parameters and measured power. As shown in Fig.~\ref{fig:forecast}, forecasting consists of training and testing phases. The training was conducted on historical data of dimension of $\mathbb{R}^{ \left \{ 94'4064 \times 5 \right \}}$, where \textcolor{black}{94'4064} is the number of minutes and 5 is the number of weather parameters. Once trained, the model predicts solar and wind power output for a \textcolor{black}{5}-minute interval given the predictions of the weather parameters for the same \textcolor{black}{5}-minute. In total, \textcolor{black}{10'656} minutes were reserved for testing purposes. A more detailed explanation of the interaction between CNN-LSTM and Wavelet Transform is given in the following steps:
\begin{enumerate}
    \item The PV and wind time series are used as an input to wavelet decomposition. The output of DWT is D1, D2, D3, and A3.
    \item The decomposed signals (D1, D2, D3, and A3) along with other input weather parameters are given as an input to predictor (CNN-LSTM) for training purpose. In this work four individual predictors are used to forecast the individual high-frequency signal (D1, D2, and D3) and low-frequency signal (A3).
    \item The output of individual predictors (D1', D2', D3', and A3') are given as an input to the reconstruction process and the output of predictor is PV power output or wind power output.
\end{enumerate}

\section{Proposed loss minimization algorithms}\label{sec:inverter}
Next, we provide a compact but sufficient description of the proposed algorithms. For more information on the two algorithms, we refer the interested reader to our previous work in \cite{arxiv_external}, where we also derive the mathematical guarantees for their performance.

Let us start with introducing the reactive power limitations of real PV inverters. First, the inverters' rated apparent power is equal to the rated active power \cite{InverterSpec}:
\begin{equation}\label{eq:SeqP}
\overline{S} = \overline{P}^G
\end{equation}

Second, the inverters can control their power factor from 0.8 over-excited to 0.8 under-excited \cite{InverterSpec}. Deriving the corresponding maximum angle $\phi^{max}$ and utilizing the relation between $cos$ and $tan$, we can express these limits via reactive and active power generations:
\begin{equation}\label{eq:PF}
-tan(\phi^{max}) \leq \frac{Q^G}{P^G} \leq tan(\phi^{max})
\end{equation}

Third, at each moment of time, the apparent power constraint should be satisfied:
\begin{equation}\label{eq:Slim}
|Q^G| \leq \sqrt{\overline{S}^2 - (P^G)^2}
\end{equation}

\begin{figure}[H]
    \centering
    \includegraphics[width=0.5\linewidth]{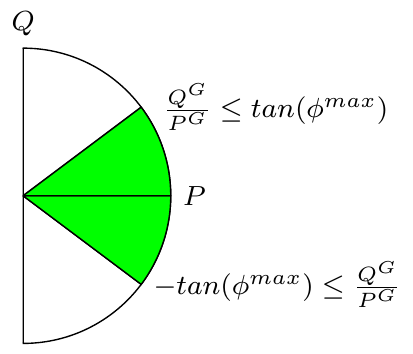}
    \caption{Phasor diagram of PV inverter.}
    \label{fig:PQDiag}
\end{figure}

The constraints (\ref{eq:PF})-(\ref{eq:Slim}) are described by the phasor diagram in Fig.~\ref{fig:PQDiag}. 

Setting reactive power setpoints to the boundary values of the constraints (\ref{eq:PF})-(\ref{eq:Slim}) implies that no reactive power reserve is left for LVRT improvement in post-contingency scenarios. 

\begin{figure}[H]
    \centering
    \includegraphics[width=0.8\linewidth]{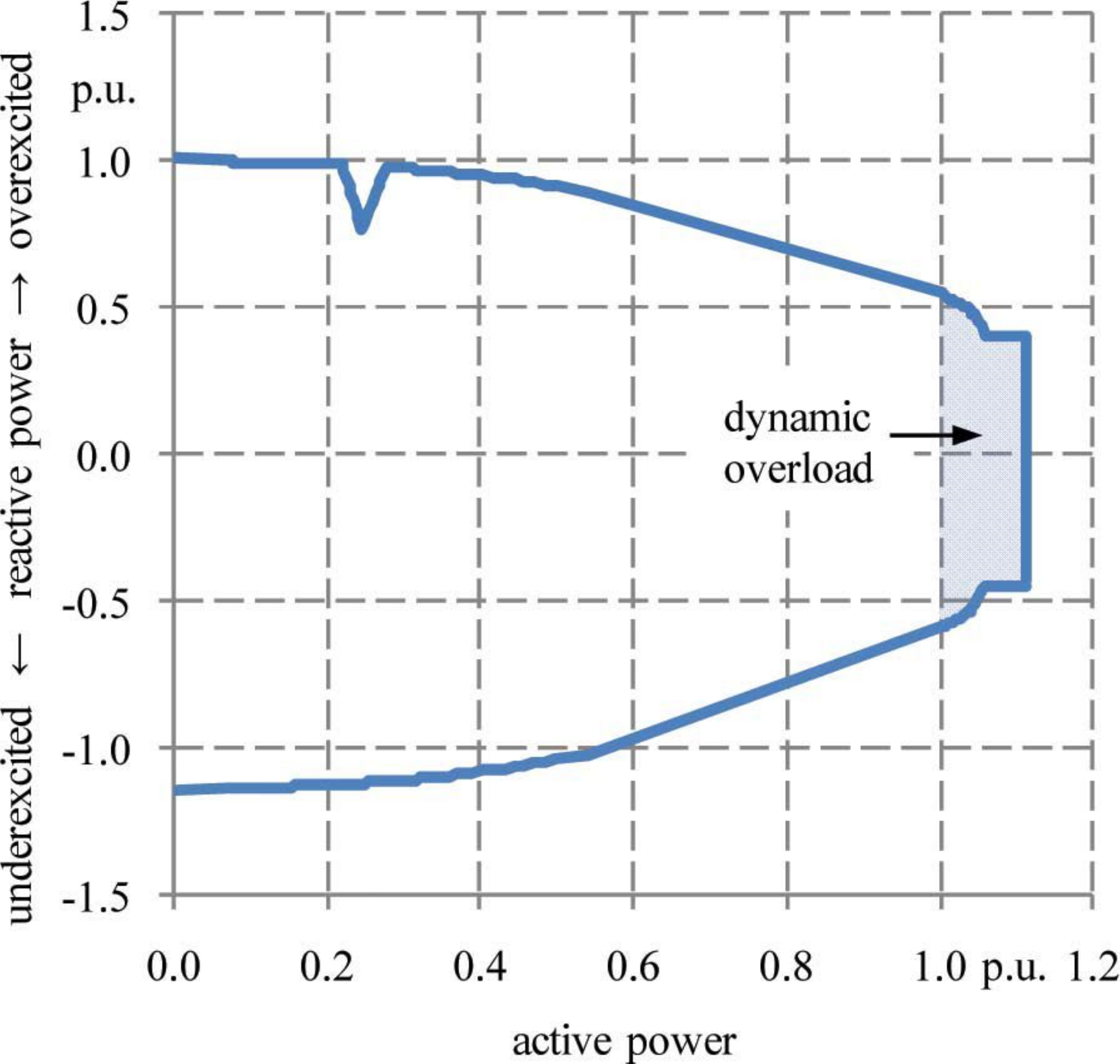}
    \caption{Reactive power capability of DFIG-WT versus active power at nominal voltage \cite{wind}.}
    \label{fig:dfig}
\end{figure}

We use the $P/Q$ capability curve from \cite{wind}, which defines the reactive power limit of a WT given the active power output at specific time, as shown in Fig.~\ref{fig:dfig}. 
Our proposed algorithms operate the obtained WT constraint similarly to PV constraints (\ref{eq:PF})-(\ref{eq:Slim}).

As shown in \cite{gur14}, reactive power injections can significantly decrease system recovery time after contingencies. As a result, we introduce reactive power reserve coefficient $k \in [0;1]$, where $k = 0$ means that no reactive power is kept as a reserve, and $k = 1$ means that all available reactive power is reserved for LVRT purposes in post-contingency cases. According to most of the grid codes, 60\% of reactive power capacity is recommended to be reserved for voltage support in contingency cases (40\% for steady state) \cite{6980139}. During numerical simulations, we test various values of the reserve coefficient $k$ (0, 0.2, 0.4, 0.6, 0.8) and derive valuable conclusions. Note that the solution enables us to dynamically adapt the reactive power reserve based on RES forecasting. More details on the motivation and procedure of dynamic reserves are given in Section \ref{sec:Theor_ExplForecast}. 

Both algorithms require only local information for their execution. As a result, we do not need to have knowledge of (or assume) the number or the location of such inverters in the system, as our algorithms are built communication-free and model-free, requiring no non-local information,  including no knowledge of the grid topology and no central coordination.

\subsection{Local Load Measuring Algorithm (LLMA)}
This algorithm is inspired by the 
\color{black}
optimization
\color{black}
approach, first proposed in \cite{c1}. We extend \cite{c1} in terms of accounting for inverter operation limits: (\ref{eq:PF})-(\ref{eq:Slim}) for PV and the $P/Q$ capability curve from \cite{wind} for WT, and call it the local load measuring algorithm (LLMA).
For LLMA setpoints, we use ${\mathcal{H}}$ superscript. For each inverter following LLMA, the only needed information is the reactive power load $Q^L$ of the same node, as shown next. We denote reactive power limits, satisfying the constraints (\ref{eq:PF})-(\ref{eq:Slim}), by $\overline{Q}^G$. LLMA is presented in Algorithm 1.
\begin{algorithm}
\caption*{\textbf{Algorithm 1: Local Load Measuring Algorithm (LLMA)}} \label{alg:LLM}
\begin{algorithmic}
    \If {$(1-k)\overline{Q}^G\geq Q^L$}
        \State $Q^{G,{\mathcal{H}}} = Q^L$
    \Else
        \State $Q^{G,{\mathcal{H}}} = (1-k)\overline{Q}^G$
    \EndIf
\end{algorithmic} 
\end{algorithm}

Considering (\ref{eq:PF}) and (\ref{eq:Slim}), Algorithm 1 can also have an equivalent form, which includes all the constraints explicitly:
\begin{multline}\label{eq:LLMA}
Q^{G,{\mathcal{H}}} = \min(Q^{L}; (1-k)P^{G} tan(\phi^{max}); \\ (1-k)\sqrt{\overline{S}^2 - (P^G)^2})
\end{multline}

\subsection{Local Flow Measuring Algorithm (LFMA)}
Next, we introduce a more advanced local algorithm, which measures the incoming flows; we call it local flow measuring algorithm (LFMA). For the final setpoints of LFMA, we use ${\mathcal{F}}$ superscript. As a prerequisite for LFMA, there is a need for reactive power flow measuring devices. LFMA consists of four steps, and each of the steps results in different settings of reactive power generation. LFMA is presented in Algorithm 2, with specified notations for generation setpoints at steps 2-4.

\begin{algorithm}
\caption*{\textbf{Algorithm 2: Local Flow Measuring Algorithm (LFMA)}} \label{alg:LFM}
\textbf{Step 1.} Branch nodes determine the upstream line by selecting a branch with the largest flow during the ``no-action'' strategy, i.e. when local reactive generation is set to 0. \\
\textbf{Step 2.} All inverters follow the same procedure as during LLMA; the reactive power generation $Q^{G,{\mathcal{H}}}$ after this step is defined by (\ref{eq:LLMA}). \\
Steps 3-4 are performed only on branch nodes, while leaf nodes do not change their own generation setpoints further.
\textbf{Step 3.} Inverters increase their own reactive generation by the value of upstream reactive flow $Q^{\mathcal{H}}_{up}$, while still satisfying the limits (\ref{eq:PF})-(\ref{eq:Slim}). The generation setpoint $Q^{G,{\mathcal{I}}}$ after step 3 is:\\
\begin{multline}\label{Alg:StepII}
Q^{G,{\mathcal{I}}} = \min(Q^{L} + Q^{\mathcal{H}}_{up}; (1-k)P^{G} tan(\phi^{max}); \\ (1-k)\sqrt{\overline{S}^2 - (P^G)^2})
\end{multline}
\textbf{Step 4.} Step 4 is performed only if upstream flow $Q_{up}$ changes direction between steps 2 and 3: 
\begin{equation}\label{Alg:Step3a}
Q^{\mathcal{H}}_{up} \cdot Q^{\mathcal{I}}_{up} < 0
\end{equation}
\begin{algorithmic}
	\If {downstream flow $Q_{do}$ does not change direction between steps 2 and 3: 
	\begin{equation}\label{Alg:Step3b}
    Q^{\mathcal{H}}_{do} \cdot Q^{\mathcal{I}}_{do} > 0
    \end{equation}}
		\State decrease reactive generation of step 3 by the absolute
		\State value of the upstream flow $|Q^{\mathcal{I}}_{up}|$: 
		\begin{equation}\label{Alg:Step3c}
        Q^{G,{\mathcal{F}}} = Q^{G,{\mathcal{I}}} - |Q^{\mathcal{I}}_{up}|
        \end{equation}
	\Else 
		    \State set reactive generation $Q^{G,{\mathcal{F}}}$ according to (\ref{eq:LLMA}).
	\EndIf
\end{algorithmic} 
\end{algorithm}

\color{black}
The interested reader is referred to \cite{arxiv_external} for illustrative examples of implementation of the proposed local algorithms on the 5-bus system.
\color{black}

\subsection{Incorporation of RES Forecast}\label{sec:Theor_ExplForecast}
We incorporate forecasting of RES to achieve greater loss decrease and improved LVRT capability. These improvements are driven by changing $Q^G$ setpoints under a forecast of $P^G$. The proposed Algorithm 3, which utilizes RES forecast, is well included in the proposed LLMA and LFMA loss minimization algorithms. 

\begin{figure}[H]
    \centering
    \includegraphics[width=0.5\linewidth]{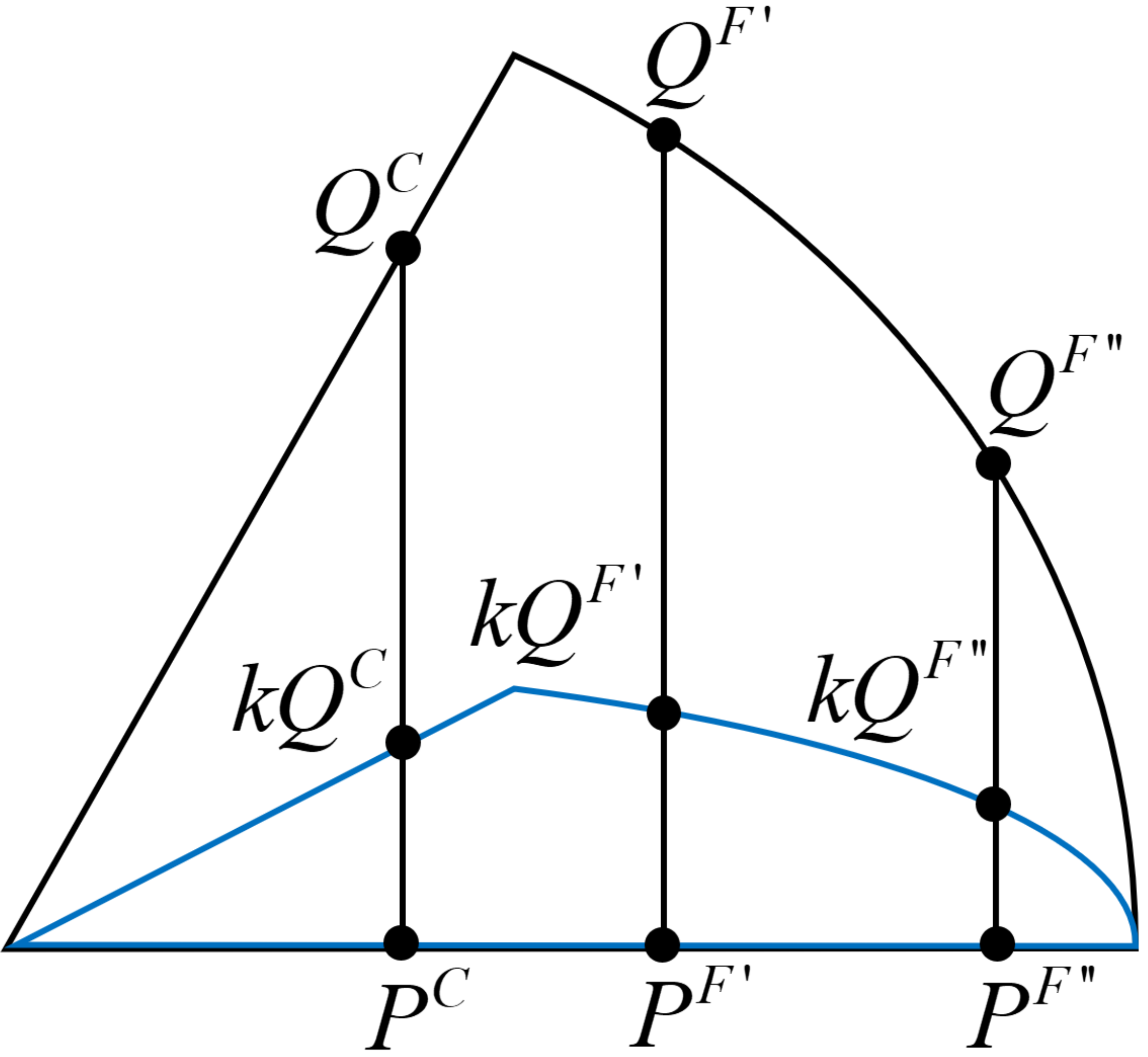}
    \caption{Incorporation of RES forecast for Algorithm 3.}
    \label{fig:RES_forecast}
\end{figure}

\begin{algorithm}
\caption*{\textbf{Algorithm 3: Incorporation of RES Forecast}} \label{alg:LLMH}
Compute $Q^F = f(P^F)$, $Q^C = f(P^C)$ by (\ref{eq:LLMA}) and/or (\ref{Alg:StepII})
\begin{algorithmic}
    \If {$Q^F > Q^C$}
        \State in $P^G = P^F$ in (\ref{eq:LLMA}) and/or (\ref{Alg:StepII})
    \Else
        \State in $P^G = P^C$ in (\ref{eq:LLMA}) and/or (\ref{Alg:StepII}) 
    \EndIf
\end{algorithmic} 
\end{algorithm}

where $P^F$ and $P^C$ are forecasted and current RES active power outputs, respectively. Note that in Algorithm 3, the value of $P^G=P^F$ is used only for computing adjusted reactive power setpoint, while active power output of RES remains set to $P^C$. 
We explain the performance of Algorithm 3 with the use of Fig.~\ref{fig:RES_forecast}, where the outer black circle segment corresponds to the upper green region of Fig.~\ref{fig:PQDiag}. The area inside the inner blue circle segment is the reactive power that is kept as a reserve for LVRT purposes, and is defined by reserve coefficient $k$. Fig.~\ref{fig:RES_forecast} shows the current active generation $P^C$ and two possible future generations $P^{F^{'}}$, $P^{F^{''}}$. Note that $Q^{F^{'}}>Q^C$, then according to Algorithm 3, $P^G=P^{F^{'}}$ in (\ref{eq:LLMA}) and$/$or (\ref{Alg:StepII}). That effectively means that more reactive power is used for loss minimization than before, i.e. without forecast. Reserve for LVRT purposes decreases, but since Algorithm 3 adjusts its setpoints every minute, and the probability of a fault in the following minute is negligibly small, the grid security is not compromised. 

\section{Numerical results}
In this section, we provide numerical results for the modified IEEE 33-bus system. The dataset to reproduce the reported results is available online \cite{code}. 
\color{black}
Note that the proposed local algorithms work for \textit{any} power distribution systems with different R/X ratio, different topologies (meshed, radial), and various equipment (switched capacitors, transformers, load tap changers) \cite{arxiv_external}. The proposed algorithm is tested on a 6-rack RTDS that is already installed in the Real Time Simulation Laboratory at the Electrical Engineering Department, Indian Institute of Technology Roorkee, India. The modeling of a modified IEEE 33-bus system with four PVs and one DFIG occupied three racks.  Due to space limitations in this paper, we provide results of RTDS simulations and MATLAB tests only for the modified IEEE 33-bus system.
\color{black}
\subsection{PV and Wind Forecasting Results}
\begin{figure}[H]
    \centering
    \vspace{-2mm}
    \includegraphics[width=0.98\linewidth]{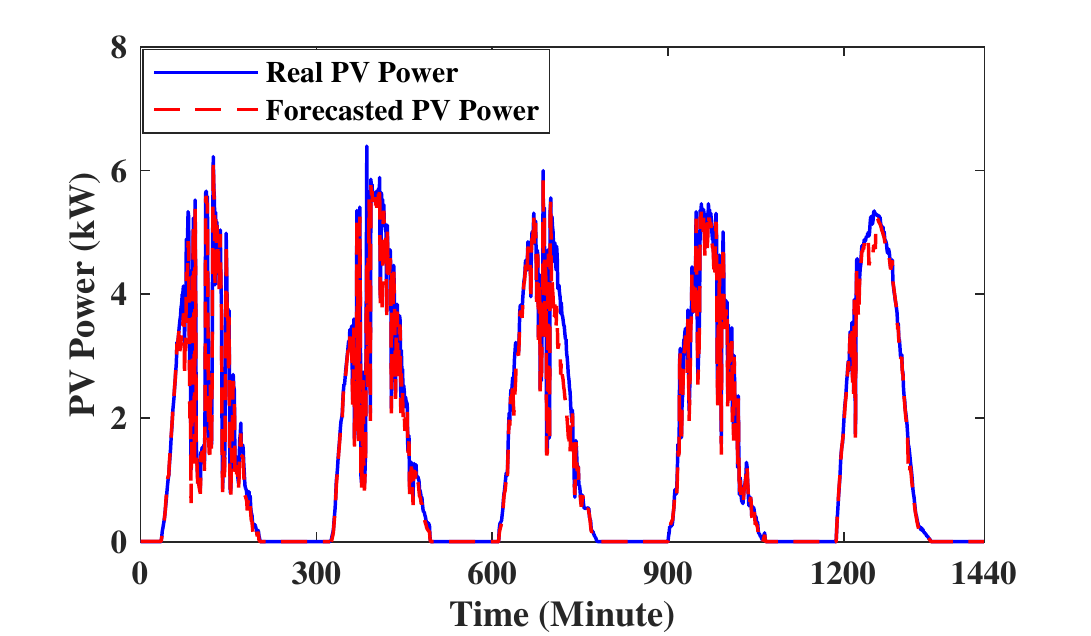} 
    \caption{PV power forecasting using Wavelet-CNN-LSTM model.}
    \label{fig:solarcnndiag}
    \vspace{-3mm}
\end{figure}

\begin{figure}[H]
    \vspace{-2mm}
    \centering
    \vspace{-2mm}
    \includegraphics [width=0.95\linewidth]{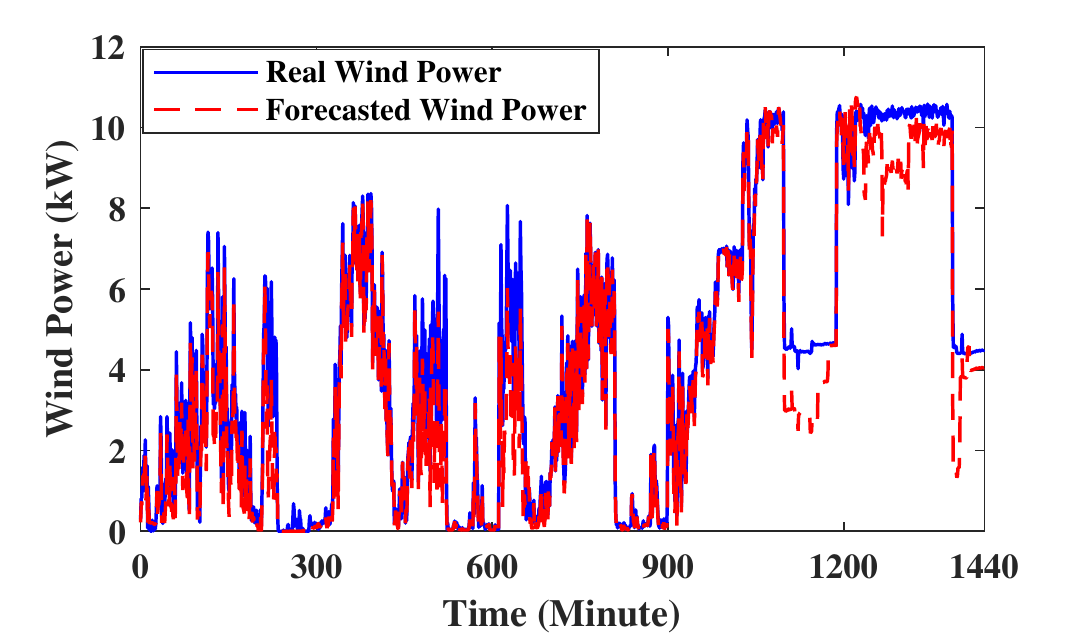} 
    \caption{Wind power forecasting using Wavelet-CNN-LSTM model.}
    \label{fig:windcnndiag}\vspace{-3mm}
\end{figure}

In this work, we implement four forecasting models: ARIMA, LSTM, CNN-LSTM, and Wavelet-CNN-LSTM. 
We focus on the mean absolute percent error (MAPE) and mean square error (MSE) metrics. The performance of each model in terms of MSE and MAPE for both PV and wind forecasting is shown Table~\ref{tab:comparison}. As shown, the Wavelet-CNN-LSTM model performs the best both for PV and wind forecasts due to the greater capturing capability of spatial and temporal properties of time series. Thus, we proceed with the Wavelet-CNN-LSTM model. PV and wind forecast results using the Wavelet-CNN-LSTM model for \textcolor{black}{1'440} out of \textcolor{black}{10'656} minutes from the testing dataset are shown in Fig.~\ref{fig:solarcnndiag} and Fig.~\ref{fig:windcnndiag}, respectively.

\begin{table}[h!]

\centering
\begin{center}
\caption{Comparison of various forecasting methods.}
\label{tab:comparison}
\begin{tabular}{lllll}
\hline\hline
                 & \multicolumn{2}{c}{PV}                                           & \multicolumn{2}{c}{Wind}                                        \\ \hline
                 & MSE        & \begin{tabular}[c]{@{}l@{}}MAPE\\ (\%)\end{tabular} & MSE       & \begin{tabular}[c]{@{}l@{}}MAPE\\ (\%)\end{tabular} \\ \hline 
ARIMA            & 0.078133 & 9.266340                                             & 0.058891  & 17.60283                                            \\ \hline
LSTM             & 0.023628 & 9.036990                                             & 0.045361  & 6.635954                                            \\ \hline
CNN-LSTM         & 0.031563 &     8.588293                                         & 0.039862  &   6.253098                                        \\ \hline
Wavelet-CNN-LSTM & 0.009207 &   7.088293                                         & 0.013768 &    5.785152                                          \\ \hline\hline
\end{tabular}
\end{center}
\end{table} 


\begin{figure*}[!h]
    \centering
    \vspace{-2mm}
    \includegraphics[width=0.86\linewidth]{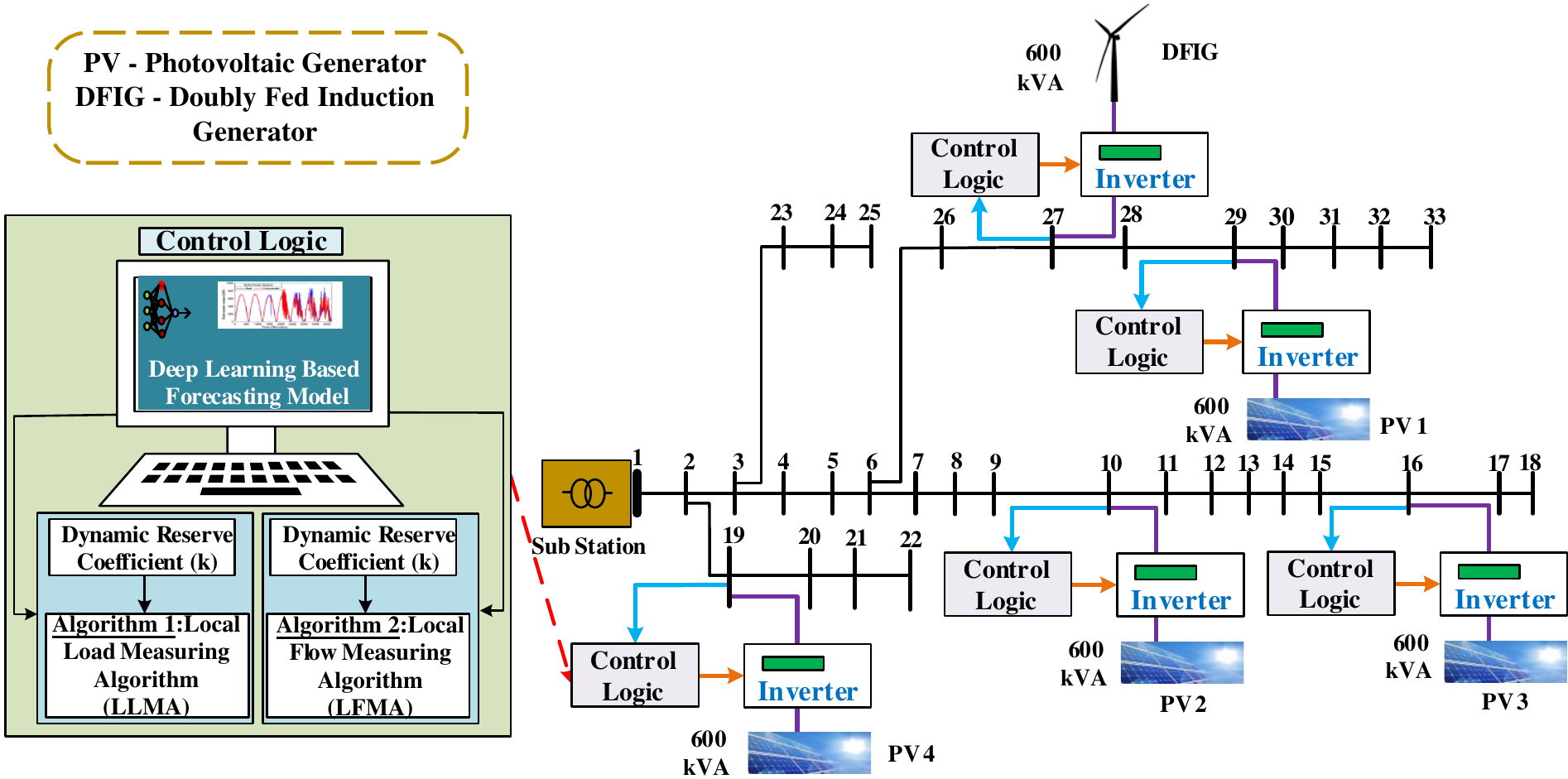} 
    \caption{Schematic diagram of applying the proposed algorithms on the modified IEEE 33-bus distribution system.}
    \label{fig:IEEE33diag}
    \vspace{-3mm}
\end{figure*}

\subsection{RTDS Simulations}
We demonstrate the capability of the proposed algorithms on the modified IEEE 33-bus system using RTDS. The schematic diagram of applying the proposed algorithms on the system is shown in Fig.~\ref{fig:IEEE33diag}. To mimic practical scenarios, containing setpoints of PVs, wind turbine, and load, a full 2019 year data from SYSLAB \cite{syslab} is used. Further, the setpoints out of the ``no-action'' algorithm, LLMA, LFMA are simulated on RTDS. During RTDS simulation, active power losses for normal operation scenarios are measured. Additionally, during the fault scenario, data of post fault voltage recovery time (PFVRT), voltage magnitude, reactive power injections are collected. 

\begin{figure}[!h]
    \centering
    \includegraphics[width=0.9\linewidth]{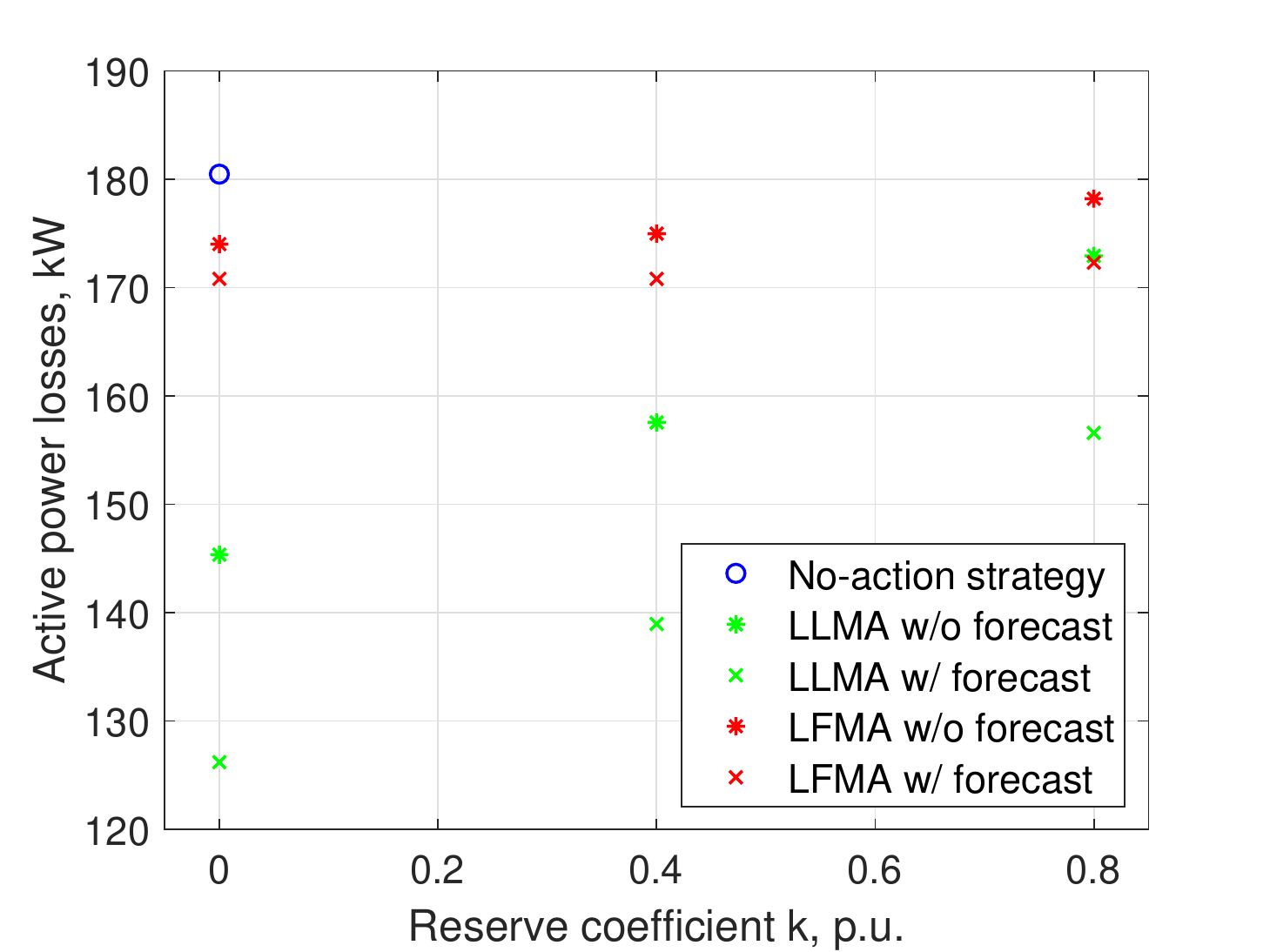}
    \caption{Comparison of loss minimization capability of the different algorithms implemented on the IEEE 33-bus system.}
    \label{fig:1}
\end{figure}

In Fig.~\ref{fig:1}, we plot active power losses for the ``no-action'' strategy, LLMA, LFMA with or without RES forecasting and under different reactive power reserve coefficients for voltage support during faults. Obviously, a higher reserve coefficient for voltage support during faults results in higher losses, since inverters have only a fraction of reactive power capacity at their disposal for loss minimization. As expected, LLMA results to lower active power losses than the ``no-action'' strategy. Moreover, LFMA, which additionally requires local information on reactive power flows, has greater loss decreasing capability and results in lower power losses than LLMA. Fig.~\ref{fig:1} confirms all of these hypotheses. Implementation of RES forecasting for LLMA and LFMA allows to additionally decrease active power losses, up to $30\%$ compared to the ``no-action'' strategy. As a result, LFMA with RES forecasting results in the lowest active power losses among all compared algorithms.

\begin{figure}[H]
    \centering
    \includegraphics[width=0.9\linewidth]{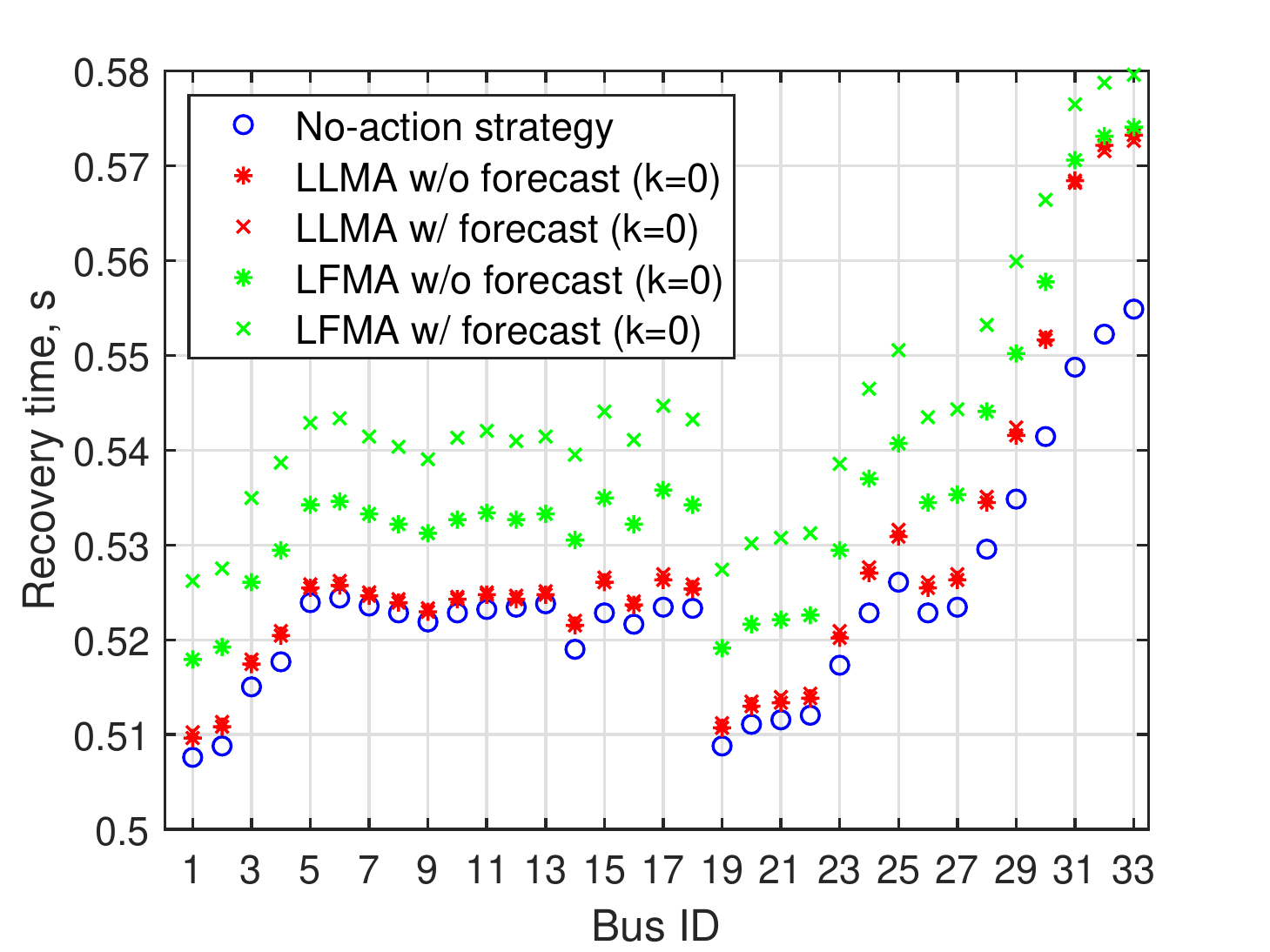}
    \caption{Comparison of recovery time of the different algorithms implemented on the IEEE 33-bus  system.}
    \label{fig:2}
\end{figure}

Next, we compare the recovery times of the algorithms. During fault, depending on the severity of voltage drop, grid codes require the injection of available reactive power to bring voltage in the acceptable range. As a result, the availability of reactive power for injection influences voltage recovery time. Considering the most extreme case $k=0$, i.e. when no reactive power is reserved for voltage support, the recovery time will be the longest. As shown in Fig.~\ref{fig:2}, for the 33-bus system, LLMA (LFMA) increases the recovery time by additional 1-20 ms (10-28 ms) depending on the bus and use of RES forecasting. This difference might be crucial if a system is at the \textit{boundary} between different operation domains. For example, semiconductor manufacturing equipment may withstand voltage $30\%$ below nominal for up to 500 ms and voltage $50\%$ below nominal for up to 200 ms \cite{semi}. If the original recovery time was close to 200 ms, an additional 28 ms can lead to the disconnection of the converter unit during very low voltages. As a result, having an option to choose the reserve coefficient value $k$ allows to establish the necessary safety margins while minimizing power losses.

\begin{figure}[!h]
    \centering
    \includegraphics[width=0.9\linewidth]{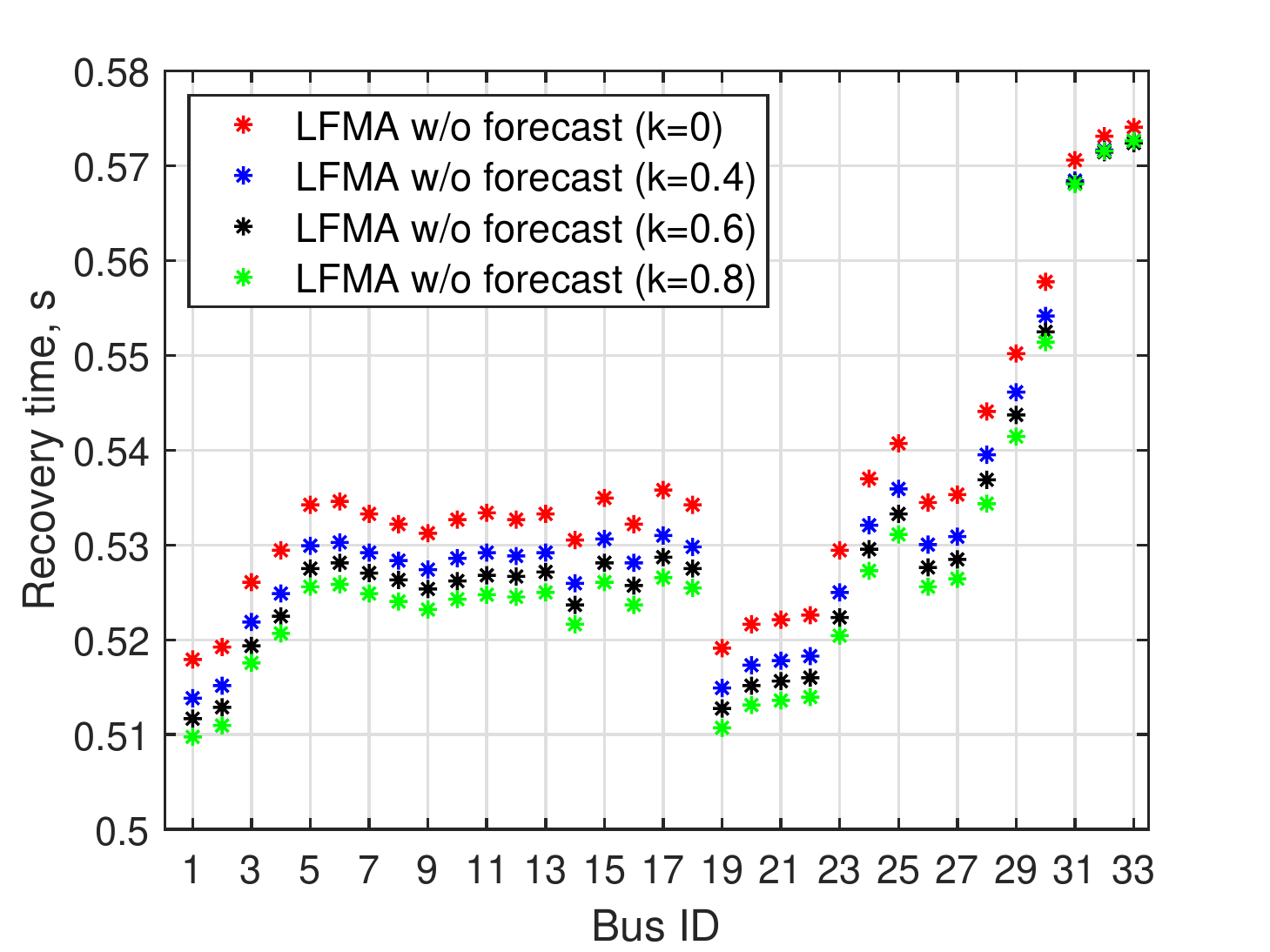}
    \caption{Comparison of recovery time of LFMA with different reserve coefficient and without forecasting capability.}
    \label{fig:14}
\end{figure}

\begin{figure}[!h]
    \centering
    \includegraphics[width=0.85\linewidth]{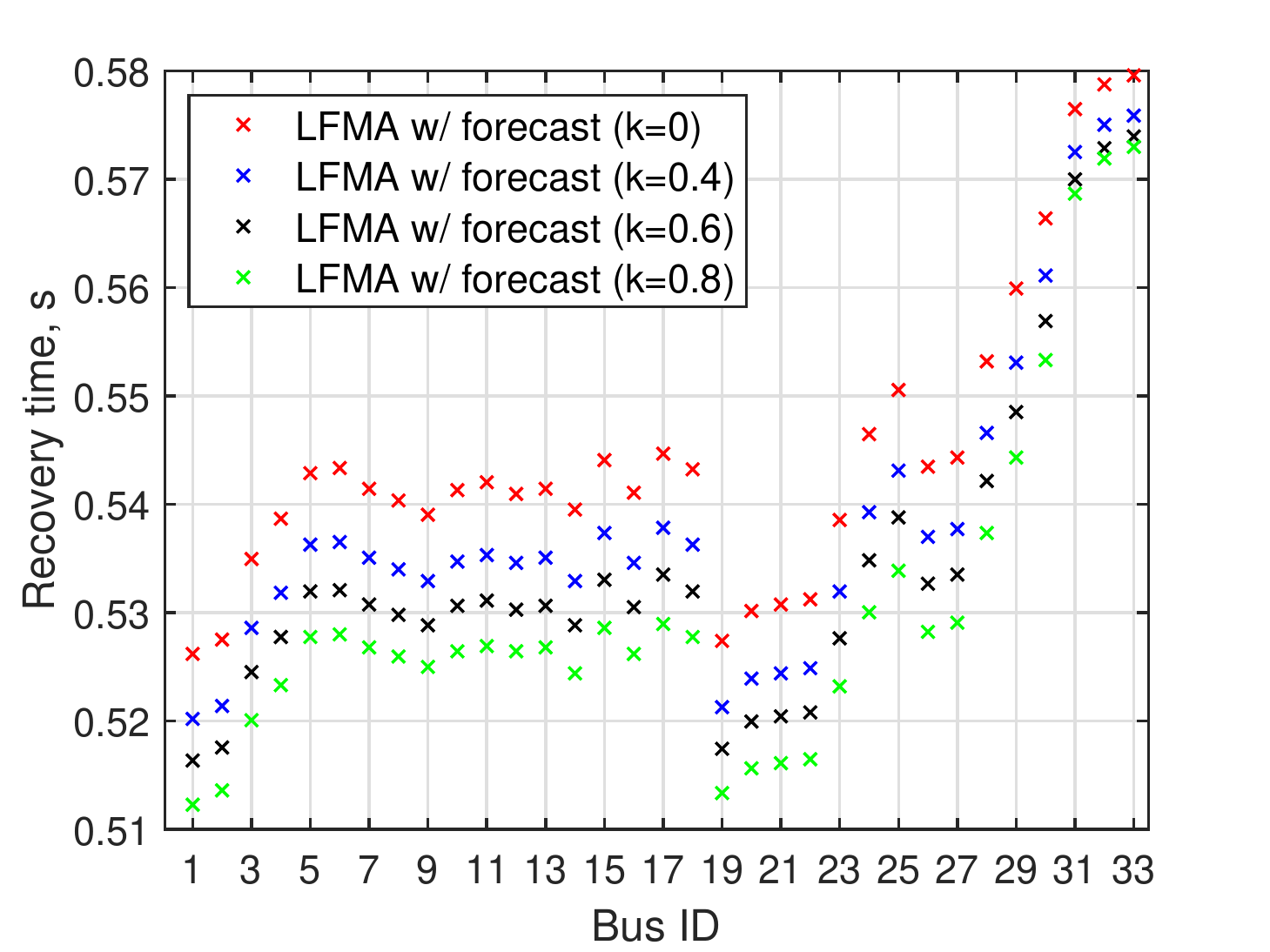}
    \caption{Comparison of recovery time of LFMA with different reserve coefficient and with forecasting capability.}
    \label{fig:15}
\end{figure}

Next, we discuss influence of reserve coefficient and RES forecasting on recovery time during LFMA operation. The obtained conclusions hold for LLMA as well. First, a greater reserve coefficient leads to shorter recovery time, which is true for both with and without RES forecast. Second, reactive power setpoints obtained with RES forecasting lead to the same or greater recovery time, since Algorithm 3 implies the same or lower reactive power reserve for voltage support. Aforementioned two points are confirmed by RTDS simulation results in Figs.~\ref{fig:14}-\ref{fig:15}. 

It is worth mentioning that recovery time for LFMA in Figs.~\ref{fig:14}-\ref{fig:15} varies between $k=0.8$ without forecasting capability and $k=0$ with forecasting capability, which for bus 17 are 0.527 s and 0.545 s, respectively. As also explained earlier in this section, for some consumers, such as semiconductor manufacturers, the difference of 18 ms in recovery time might be crucial, so the choice of $k$ and enabling or not forecasting capability are important questions \cite{semi}. In that case, our proposed algorithms provide a flexible toolkit for choosing both of these parameters in order to minimize active power losses while keeping recovery time according to the specific grid code. 

\begin{figure}[H]
    \centering
    \includegraphics[width=1.02\linewidth]{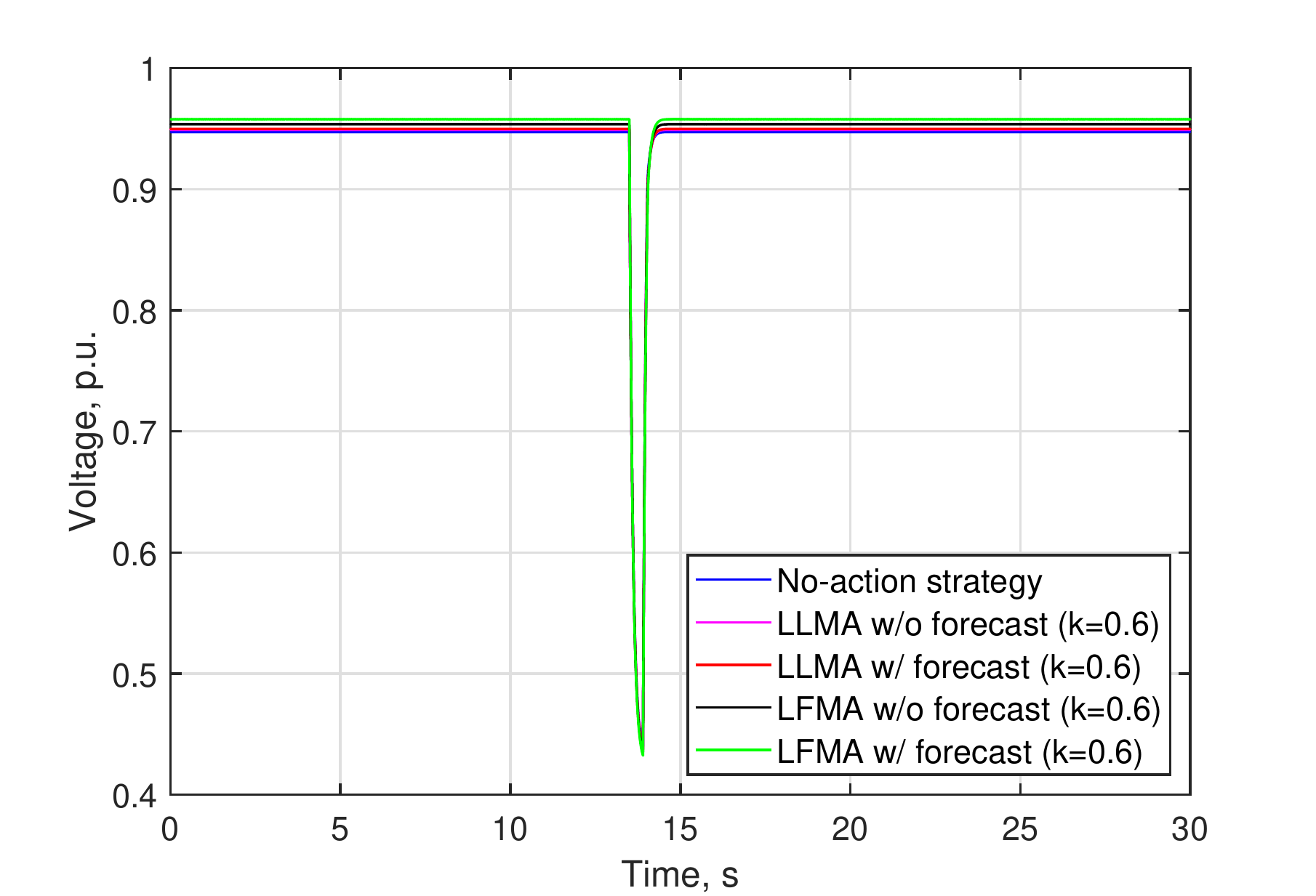}
    \caption{Voltage magnitudes on a bus 27, equipped with DFIG, under the fault event and setpoints of the different algorithms for $k=0.6$.}
    \label{fig:6}
\end{figure}

During normal operation, LLMA and LFMA keep the voltages at all buses within the permitted limits, which are [0.9; 1.1] p.u. in this work. Fig.~\ref{fig:6} presents the voltages on a bus 27 before, during, and after a fault under the control of different algorithms for $k=0.6$. Note that the fault occurs at around 13.48 s, which results in a voltage drop below 0.5 p.u.. According to the grid code, in this case, all available reactive power should be injected in order to support the voltage. As a result, by 14.48 s the voltage at all buses is restored to the pre-fault values. 

\begin{figure}[!h]
    \centering
    \includegraphics[width=1.1\linewidth]{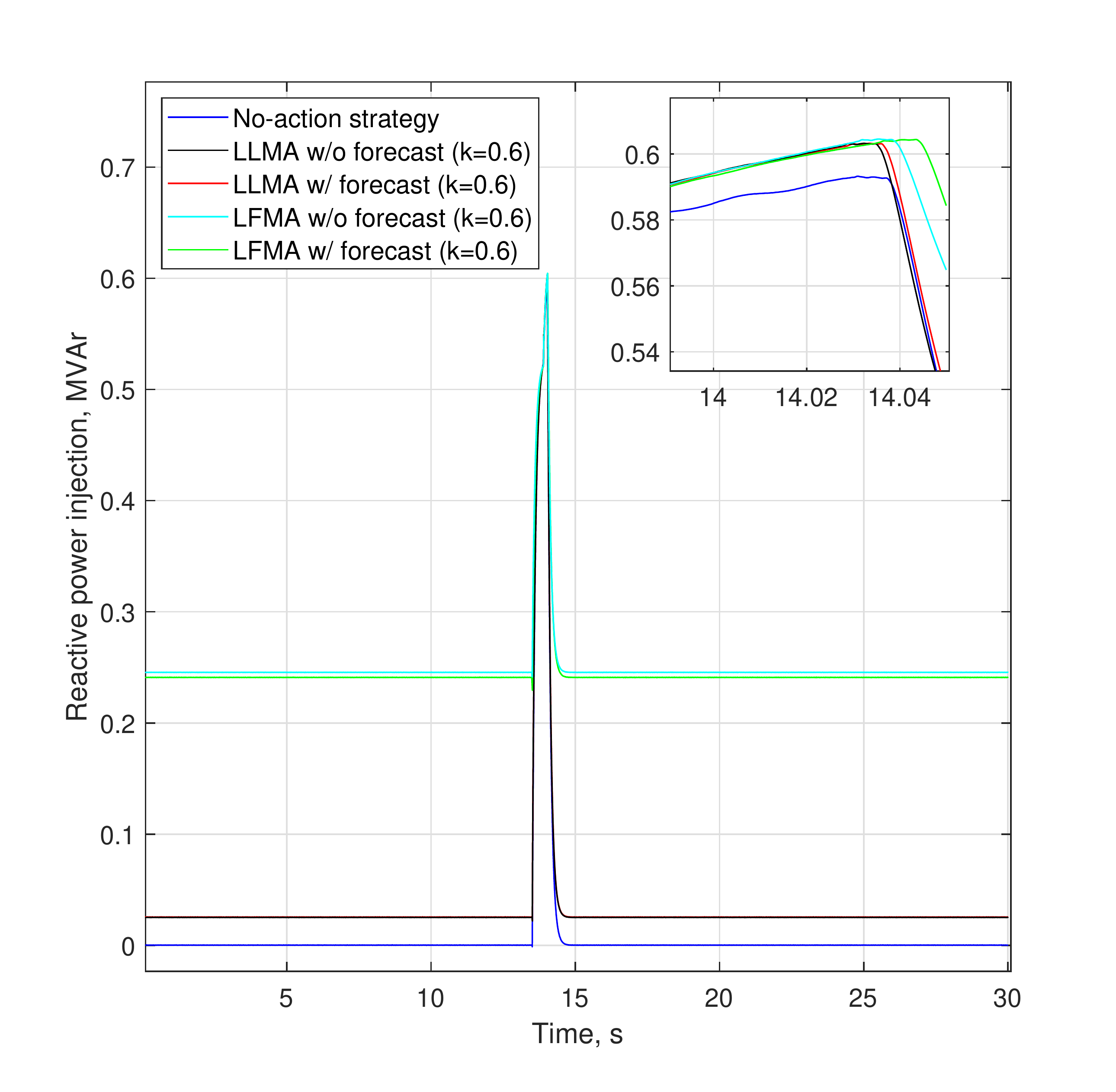}
    \caption{Reactive power setpoints of DFIG on a bus 27 under the fault event and execution of the different algorithms.}
    \label{fig:7}
\end{figure}

The amount of reactive power additionally injected during the fault depends on the maximum reactive power capability of an inverter and its pre-fault reactive power setpoint. Obviously, the ``no-action'' strategy, LLMA, LFMA result in different pre- and post-fault reactive power setpoints for any $k\neq1$. During normal operation, the ``no-action'' strategy injects zero reactive power, LLMA injects more than the ``no-action'' strategy but less than LFMA. Also, during normal operation, both LLMA and LFMA with RES forecasting inject equal or more than their versions without forecasting. All of these statements are confirmed by RTDS simulation results in Fig.~\ref{fig:7}. Observe that since the voltage falls below 0.5 p.u. during the fault, all algorithms inject the maximum available reactive power capacity, which is 0.6 MVar for the bus 27. As shown in the enlarged subplot in Fig.~\ref{fig:7}, the algorithms are almost identical in injecting the maximum reactive power capacity.

\begin{table*}[!t]
\centering
\begin{center}
\caption{Comparison of the communication-based and communication-free algorithms in the IEEE 33-bus system with PV inverters grid-connected and grid-disconnected at night hours for one day of SYSLAB data simulations in MATLAB.}
\label{tab:NightHour}
\begin{tabular}{p{0.15\textwidth}<{\centering}p{0.22\textwidth}<{\centering\arraybackslash}p{0.14\textwidth}<{\centering}p{0.01\textwidth}<{\centering}p{0.22\textwidth}<{\centering}p{0.14\textwidth}<{\centering}}
\hline
Algorithm & \multicolumn{2}{c}{Grid-connected at night hours} & & \multicolumn{2}{c}{Grid-disconnected at night hours} \\ \cline{2-3} \cline{5-6} 
    &  Average active power losses, (kW) & Energy losses, (kWh) &  & Average active power losses, (kW) & Energy losses, (kWh) \\ \hline
``No-action'' strategy & 9.66 & 231.52 & & 9.73 & 233.21 \\ 
LLMA & 9.42 & 225.82  & & 9.53 & 228.58 \\ 
LFMA & 7.59 & 182.09  & & 7.69 & 184.51 \\ 
Centralized OPF & 8.95 & 214.71  & & 9.39 & 225.21 \\ \hline
\end{tabular}
\end{center}
\end{table*}

\subsection{MATLAB Simulations}
RTDS simulations are ideal for simulating with high precision the behavior of a real system during faults. However, the simulation timeline on RTDS is limited to minutes or hours, at maximum. As a result, to model the performance of the proposed algorithms on extended time intervals, MATLAB simulations are carried out and MATPOWER 7 is used for computing the AC power flows. The full day simulations are conducted for 20 August 2019, and the corresponding time series are extracted from SYSLAB data \cite{syslab}. Further, the night hours are defined as the time between sunset and sunrise. For SYSLAB facilities in Denmark on 20 August 2019, the night hours are between 20:00-05:00, during which PV inverters are either connected to the grid or not. While in most countries the PV inverters are grid-disconnected at night, the results in Table~\ref{tab:NightHour} show that keeping them connected would provide around $4-15\%$ decrease in active power and energy losses for the considered algorithms. Note that values for the ``no-action'' strategy differ for grid-connected and grid-disconnected cases since there is still small PV generation during the defined night hours. New reactive power setpoints are computed each minute. 
Additionally, we provide results for a centralized optimal power flow (OPF) setup, where the perfect system model is known and fast communication is available, which is usually impossible for distribution systems. If the centralized OPF does not converge, i.e. is unable to solve the optimization problem due to limitations of the non-linear solver, then the ``no-action'' strategy is performed, as we assume that communication-free algorithms are not established in that case. Observe that for both grid-connected and grid-disconnected cases, LFMA outperforms centralized OPF by achieving lower power and energy losses. Such robust performance of LFMA while utilizing only local information is a definite advantage over centralized OPF. 

\section{Conclusion}
Algorithms designed to minimize losses through the adjustment of the converter reactive power setpoints often neglect the impact they have on the fault-ride-through capability of the converters. This paper proposes algorithms that consider both: they achieve a substantial decrease in power losses, while their impact on the recovery time after a fault is up to a few tens of milliseconds. This ensures that the distributed generation will remain connected to the grid and assist towards its recovery. Our proposed algorithms are decentralized and model-free: they require no communication and no knowledge of the grid topology or the grid location of the converters. Through extensive RTDS simulations, we show how our proposed algorithms have the ability to adjust their reactive power reserves in real-time and contribute to reduced voltage recovery times during faults. Running these algorithms under normal conditions for a full day, we also highlight the benefits of the communication-free and model-free approaches we propose in this paper: while the non-linear OPF solver does not always converge to a feasible solution due to its complexity, our proposed algorithm can always determine a feasible solution which at times achieves energy losses that are even lower than the centralized OPF, despite not having full knowledge and full control over the system.

\bibliographystyle{IEEEtran}
\bibliography{library}

\end{document}